\documentclass[english]{article}
\usepackage[T1]{fontenc}
\usepackage{color}
\usepackage{graphicx}
\usepackage{amssymb}

\makeatletter

\usepackage{a4}
\input{epsfig.sty}
\def\fnum@table{\tablename~{\bf\thetable}}
\def\fnum@figure{\figurename~{\bf\thefigure}}
\def\tablename{\footnotesize{\bf Table}}
\def\figurename{\footnotesize{\bf Figure}}

\def\be{\begin{equation}}
\def\ee{\end{equation}}

\usepackage{babel}
\makeatother
\begin{document}

\title{\textbf{On the re-summation of enhanced Pomeron diagrams}}

\author{\textbf{S. Ostapchenko}%
\footnote{e-mail: serguei@ik.fzk.de%
} \textbf{}\\
\textit{\small Forschungszentrum Karlsruhe, Institut} \textit{\textcolor{black}{\small für}}
\textit{\small Kernphysik, 76021 Karlsruhe, Germany}\\
\textit{\small D.V. Skobeltsyn Institute of Nuclear Physics, Moscow
State University, 119992 Moscow, Russia} \textit{}\\
}

\maketitle
\begin{center}\textbf{\large Abstract}\end{center}{\large \par}

Dominant contributions of enhanced Pomeron diagrams to elastic hadron-hadron
scattering amplitude are re-summed to all orders. The formalism is
applied to calculate total hadronic cross sections and elastic scattering
slopes. An agreement with earlier results is obtained.

\section{Introduction\label{intro.sec} }

\begin{figure}[hbt]
\begin{center}\includegraphics[%
  width=10cm,
  height=4cm]{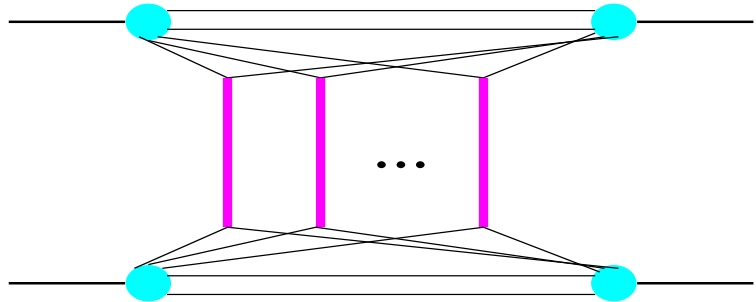}\end{center}
\vspace{-4mm}

\caption{General multi-Pomeron contribution to hadron-hadron scattering amplitude;
elementary scattering processes (vertical thick lines) are described
as Pomeron exchanges.\label{multiple}}
\end{figure}

Despite a significant progress in the perturbative QCD during last
decades one still has to rely on phenomenological approaches when
calculating total hadron-hadron cross sections, diffraction dissociation
probabilities, or when treating particle production in general minimum
bias hadronic collisions. The most powerful one proved to be the Gribov's
Reggeon approach \cite{gri68}, where high energy interactions
are described as multiple scattering processes, elementary re-scatterings
being treated phenomenologically as Pomeron exchanges, as shown in
Fig.~\ref{multiple}.%

Usual ansatz for the Pomeron amplitude in impact parameter representation
is \cite{bak76}
\begin{eqnarray}
f_{ad}^{\mathbb{P}}(s,b)=\frac{i\,\gamma_{a}\gamma_{d}\,
(s/s_{0})^{\alpha_{\mathbb{P}}(0)-1}}{\lambda_{ad}(s)}
\,e^{-\frac{b^{2}}{4\lambda_{ad}(s)}}\label{ampl-P}\\
\lambda_{ad}(s)=R_{a}^{2}+R_{d}^{2}+\alpha_{\mathbb{P}}'(0)\,\ln(s/s_{0}),\label{lambda-P}\end{eqnarray}
where $s$ and $b$ are c.m.~energy squared and impact parameter
for the interaction, $s_{0}\simeq1$ GeV$^{2}$ is the hadronic mass
scale, $\alpha_{\mathbb{P}}(0)$ and $\alpha_{\mathbb{P}}'(0)$ are
the intercept and the slope of the Pomeron Regge trajectory, and $\gamma_{a}$,
$R_{a}^{2}$ are the coupling and the slope of Pomeron-hadron $a$
vertex.

Considering any number of Pomerons exchanged between hadrons $a$
and $d$ one obtains elastic scattering amplitude as a sum of contributions
of diagrams of Fig.~\ref{multiple} \cite{bak76}:
\begin{eqnarray}
i\, f_{ad}(s,b)=\frac{1}{C_{a}\, C_{d}}\sum_{n=1}^{\infty}
\frac{\left[i\, C_{a}\, C_{d}\, f_{ad}^{\mathbb{P}}(s,b)\right]^{n}}{n!}
=\frac{1}{C_{a}\, C_{d}}\,\left[e^{-\chi_{ad}^{\mathbb{P}}(s,b)}-1\right]
\label{f_ad}\\
\chi_{ad}^{\mathbb{P}}(s,b)=
\frac{1}{i}\, C_{a}\, C_{d}\, f_{ad}^{\mathbb{P}}(s,b),
\label{P-eik}\end{eqnarray}
where shower enhancement coefficient $C_{a}$ accounts for low mass
inelastic intermediate states for hadron $a$ and $\chi_{ad}^{\mathbb{P}}$
is the Pomeron quasi-eikonal.

This allows one to calculate total cross section $\sigma_{ad}^{{\rm tot}}$
and elastic scattering slope $B_{ad}^{{\rm el}}$ as 
\begin{eqnarray}
\sigma_{ad}^{{\rm tot}}(s)=2\,{\rm Im}\int\!\! d^{2}b\, f_{ad}(s,b)
=\frac{2}{C_{a}\, C_{d}}\int\!\! d^{2}b\left[1-e^{-\chi_{ad}^{\mathbb{P}}(s,b)}\right]\label{sigma-tot}\\
B_{ad}^{{\rm el}}(s)=\left.\frac{d}{dt}\ln\frac{d\sigma_{ad}^{{\rm el}}(s,t)}
{dt}\right|_{t=0}
=\frac{1}{C_{a}\, C_{d}\,\sigma_{ad}^{{\rm tot}}(s)}\int\!\! d^{2}b\; b^{2}\left[1-e^{-\chi_{ad}^{\mathbb{P}}(s,b)}\right],\label{b_el}\end{eqnarray}
where $d\sigma_{ad}^{{\rm el}}(s,t)/dt$ is the differential elastic
cross section for momentum transfer squared $t$. Comparing to data
one obtains typically $\alpha_{\mathbb{P}}(0)\simeq1.1$,
 $\alpha_{\mathbb{P}}'(0)\simeq0.2$ GeV$^{-2}$ \cite{kai82}.

Making use of Abramovskii-Gribov-Kancheli  cutting rules \cite{agk}
allows one to re-sum contributions of various unitarity cuts of the
elastic scattering diagrams of Fig.~\ref{multiple}, corresponding
to particular inelastic final states of hadron-hadron interactions.
This opens the way for developing powerful model approaches \cite{kai82}
and for constructing Monte Carlo generators for hadronic and
nuclear collisions \cite{aur92}.

Still, the underlying picture for the described scheme corresponds
to an interaction mediated by a number of independent parton cascades.
In {}``dense'' regime, i.e.~in the limit of high energies and small
impact parameters, one expects a large number of such elementary scattering
processes. Then,  the  underlying parton cascades should largely overlap
and interact with each other, giving rise to significant non-linear
effects \cite{glr}. The latter are traditionally described by enhanced
Pomeron diagrams, which involve Pomeron-Pomeron interactions 
\cite{kan73,car74}.
However, consistent treatment of enhanced corrections proved to
be a very non-trivial problem: as the energy increases more and more
diagrams of complicated topologies come into play. 

General approach to the re-summation of higher order enhanced graphs 
has been proposed
in \cite{kai86} assuming $\pi$-meson dominance of multi-Pomeron
vertices. It has been shown that in the limit of very high energies
the full Pomeron scheme is equivalent to the above-described quasi-eikonal
picture, however, based on a Pomeron with suitably renormalized intercept.
An alternative procedure was suggested in \cite{bon01}, where one
re-summed dominant enhanced corrections to hadron-nucleus and nucleus-nucleus
scattering amplitudes postulating a negligibly small Pomeron slope
and including only triple-Pomeron vertices. However,  corresponding 
algorithms can not be applied to treat  general
 inelastic final states in hadronic interactions.

Present work is devoted to a direct all-order
 re-summation of enhanced contributions
to hadron-hadron elastic scattering amplitude. The analysis of corresponding
unitarity cuts and applications of the approach to particle production
treatment are discussed elsewhere \cite{ost06}.

\section{The formalism\label{lin-scheme}}

Comparatively simple generalization of the previous scheme can be
obtained using  eikonal vertexes $g_{mn}$ for the
transition of $m$ into $n$ Pomerons, 
$g_{mn}=r_{3\mathbb{P}}\,\gamma_{\mathbb{P}}^{m+n-3}/(4\pi \,m!\,n!)$,
with $r_{3\mathbb{P}}$ being the triple-Pomeron coupling. In particular,
assuming pion dominance of multi-Pomeron vertices, 
one has $g_{mn}=\frac{G}{C_{\pi}^{2}}\,(C_{\pi}\,\gamma_{\pi})^{m+n}/m!/n!$
and the vertex slope $R_{\mathbb{P}}^{2}$ is equal to the one of
Pomeron-pion coupling $R_{\pi}^{2}$ \cite{car74,kai86}. Thus, for
a Pomeron exchanged between two vertices one uses the quasi-eikonal
for pion-pion scattering $\chi_{\pi\pi}^{\mathbb{P}}(s_{0}\,e^{\Delta y},\Delta b)$,
as defined by (\ref{P-eik}), (\ref{ampl-P}-\ref{lambda-P}), where
$\Delta y$ and $\Delta b$ are rapidity and impact parameter
distances between the vertices. Correspondingly, a Pomeron exchange
between hadron $a$ and a given vertex is described by 
$\chi_{a\pi}^{\mathbb{P}}(s_{0}\,e^{\Delta y},\Delta b)$.
This way the  contribution with only one multi-Pomeron
vertex is obtained using standard Reggeon calculus techniques
\cite{gri68,bak76}: summing over $m\geq1$ Pomerons exchanged
between the vertex and the projectile hadron, $n\geq1$ Pomeron exchanges
between the vertex and the target, subtracting the term with $m=n=1$
(Pomeron self-coupling), and integrating over rapidity 
$y_{1}<Y=\ln\frac{s}{s_{0}}$ and
impact parameter $\vec{b}_{1}$ of the vertex \cite{kai86},
see Fig.~\ref{PPP-1}:%
\begin{figure}[t]
\begin{center}\includegraphics[%
  width=6cm,
  height=4cm]{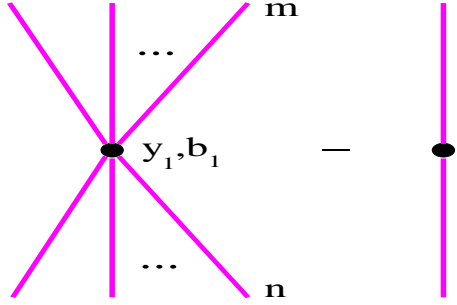}\end{center}

\vspace{-4mm}

\caption{Lowest order enhanced graphs; Pomeron connections to the projectile
and target hadrons not shown explicitely.\label{PPP-1}}
\end{figure}%
\begin{eqnarray}
\Delta\chi_{ad}^{\mathbb{PPP}}\!(s,b)=
\frac{G}{C_{\pi}^{2}}\sum_{m,n\geq1;m+n\geq3}
\int_{0}^{Y}\!\!\! dy_{1}\;
\int\!\! d^{2}b_{1}\;
\frac{\left[-\chi_{a\pi}^{\mathbb{P}}(s_0\,e^{Y-y_{1}},
|\vec{b}-\vec{b}_{1}|)\right]^{m}}{m!}\;
\frac{\left[-\chi_{d\pi}^{\mathbb{P}}(s_{0}\,e^{y_{1}},b_{1})\right]^{n}}{n!}
\nonumber \\ 
=\frac{G}{C_{\pi}^{2}}\int_{0}^{Y}\!\!\! dy_{1}
\!\int\!\! d^{2}b_{1}\;
 \left\{\left(1-e^{-\chi_{a\pi}^{\mathbb{P}}
(s_0\,e^{Y-y_{1}},|\vec{b}-\vec{b}_{1}|)}\right)\,
\left(1-e^{-\chi_{d\pi}^{\mathbb{P}}(s_{0}\,e^{y_{1}},b_{1})}
\right)\right.
\nonumber \\
\left.-\chi_{a\pi}^{\mathbb{P}}(s_0\,e^{Y-y_{1}},|\vec{b}-\vec{b}_{1}|)
\;\chi_{d\pi}^{\mathbb{P}}(s_{0}\,e^{y_{1}},b_{1})\right\} 
\label{3P-(1)}\end{eqnarray}

At $s\rightarrow\infty$ and $b,b_{1}\rightarrow0$ the integrand
in the r.h.s.~of (\ref{3P-(1)}) is dominated by the last term in
the curly brackets, which corresponds to the contribution of the subtracted
graph in Fig.~\ref{PPP-1}. Thus, assuming that the main contribution
to the integral over $\vec{b}_{1}$ in (\ref{3P-(1)}) comes from
comparatively small $b_{1}$ and neglecting the slope of the multi-Pomeron
vertex, one obtains asymptotically 
\begin{eqnarray}
\Delta\chi_{ad}^{{\rm asymp}(1)}(s,b)\sim-4\pi\, G\,\gamma_{\pi}^{2}\,
\ln\!\frac{s}{s_{0}}\,\chi_{ad}^{\mathbb{P}}(s,b)&&
\label{asy-(1)}\end{eqnarray}

Under the above assumptions the contributions of higher order graphs also
reduce in the {}``dense'' limit to subtracted Pomeron self-couplings,
which leads to \cite{kai86}
\begin{eqnarray}
\Delta\chi_{ad}^{{\rm asymp}}(s,b)\sim
\left[(s/s_{0})^{-4\pi G\gamma_{\pi}^{2}}-1\right]
\chi_{ad}^{\mathbb{P}}(s,b)&&
\label{asy-tot}\end{eqnarray}

Thus, asymptotically one recovers the usual quasi-eikonal scheme based
on re-normalized quasi-eikonal 
$\tilde{\chi}_{ad}^{\mathbb{P}}=
\chi_{ad}^{\mathbb{P}}+\Delta\chi_{ad}^{{\rm asymp}}$, which is
 defined by (\ref{P-eik}), (\ref{ampl-P}-\ref{lambda-P})
with the Pomeron intercept
$\tilde{\alpha}_{\mathbb{P}}(0)=\alpha_{\mathbb{P}}(0)
-4\pi\, G\,\gamma_{\pi}^{2}$.

Our goal is to re-sum dominant enhanced diagrams to all orders, to
obtain a smooth transition between the {}``dilute'' (small energies,
large impact parameters) and {}``dense'' regimes. As in \cite{kai86},
we assume pion dominance of multi-Pomeron vertices; treating
$r_{3\mathbb{P}}=4\pi \,G\,C_{\pi}\gamma_{\pi}^3$, 
$\gamma_{\mathbb{P}}=C_{\pi}\gamma_{\pi}$, $R_{\mathbb{P}}^{2}=R_{\pi}^{2}$ 
as free parameters
one may recover the  general eikonal form for the Pomeron vertices $g_{mn}$
in case of proton-proton scattering.

\begin{figure}[htb]
\begin{center}\includegraphics[%
  width=8cm,
  height=4cm]{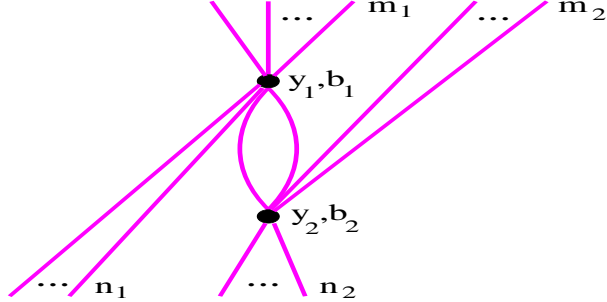}\end{center}

\vspace{-4mm}

\caption{An example of a {}``loop'' graph.\label{loop}}
\end{figure}

We start from recalling that one can neglect contributions of {}``loop''
graphs, which contain multi-Pomeron vertices connected to each other
by at least two Pomerons, as shown in Fig.~\ref{loop}.%
 Indeed, in
the {}``dilute'' regime such contributions are suppressed by powers
of small triple Pomeron coupling $G$ ($G^{2}$ for the diagram of
Fig.~\ref{loop}). On the other hand, in the {}``dense'' limit,
summing over any number $n_{1}$ of Pomerons exchanged between the
target and the upper vertex in Fig.~\ref{loop} (similarly for the
lower vertex), one obtains an exponential factor which strongly dumps
the overall contribution \cite{car74,kai86}:
\begin{eqnarray}
\sum_{n_{1}=0}^{\infty}\frac{\left[-\chi_{d\pi}^{\mathbb{P}}
(s_{0}\,e^{y_{1}},b_{1})\right]^{n_{1}}}{n_{1}!}=
e^{-\chi_{d\pi}^{\mathbb{P}}(s_{0}\,e^{y_{1}},b_{1})}&&\label{dump}
\end{eqnarray}

\begin{figure}[htb]
\begin{center}\includegraphics[%
  width=10cm,
  height=4cm]{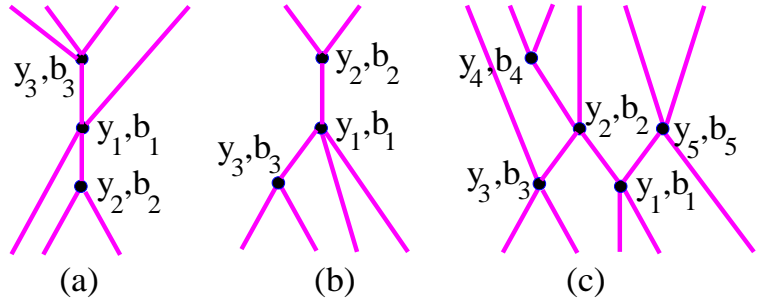}\end{center}
\vspace{-4mm}

\caption{Enhanced Pomeron graphs of {}``tree'' type (a,b) and of {}``net''
type (c).\label{net}}
\end{figure}

Let us first re-sum contributions of {}``tree''-type graphs, shown
in Fig.~\ref{net}a,b.%
 Those contain one {}``central'' (not necessarily unique, see Fig.~\ref{net}b)
vertex, from which a number of Pomeron {}``fans'' develops towards
the projectile and the target.

General {}``fan'' contribution can be defined via the recursive
equation of Fig.~\ref{ffan}:%
\begin{figure}[htb]
\begin{center}\includegraphics[%
  width=10cm,
  height=4cm]{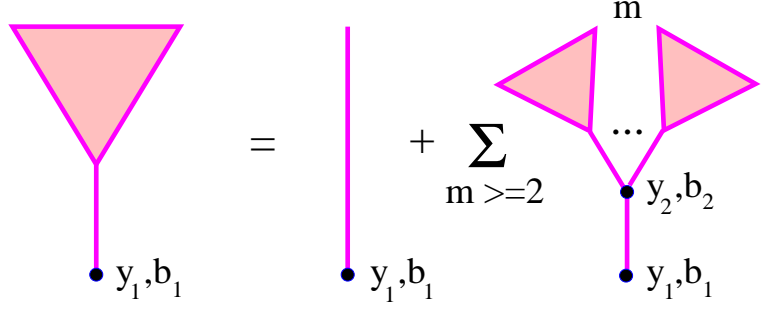}\end{center}

\vspace{-4mm}

\caption{Recursive equation for the {}``fan'' contribution $\chi_{a}^{{\rm fan}}(y_{1},b_{1})$;
$y_{1}$ and $b_{1}$ are rapidity and impact parameter distances
between hadron $a$ and the vertex in the {}``handle'' of the {}``fan''.\label{ffan}}
\end{figure}
\begin{eqnarray}
\chi_{a}^{{\rm fan}}(y_{1},b_{1})=
\chi_{a\pi}^{\mathbb{P}}(s_{0}\,e^{y_{1}},b_{1})+\frac{G}{C_{\pi}^{2}}
\int_{0}^{y_{1}}\!\! dy_{2}\int\!\! d^{2}b_{2}\;\;
\chi_{\pi\pi}^{\mathbb{P}}(s_{0}\,e^{y_{1}-y_{2}},|\vec{b}_{1}-\vec{b}_{2}|)
\nonumber \\
\times\left[1-e^{-\chi_{a}^{{\rm fan}}(y_{2},b_{2})}
-\chi_{a}^{{\rm fan}}(y_{2},b_{2})\right]\label{fan}\end{eqnarray}

\begin{figure}[htb]
\begin{center}\includegraphics[%
  width=10cm,
  height=4cm]{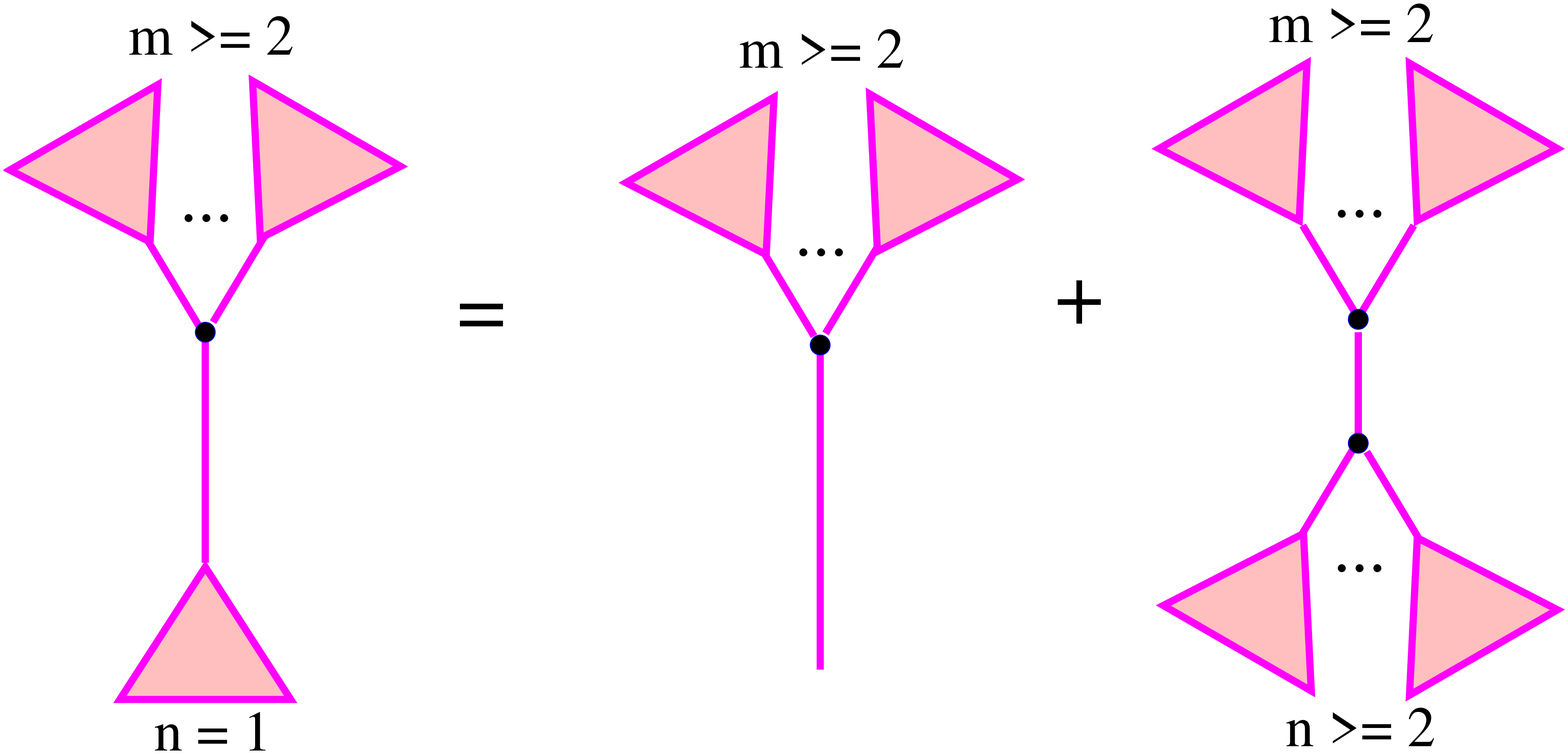}\end{center}
\vspace{-4mm}

\caption{{}``Tree'' graphs with only one target {}``fan'' can be expanded
as shown in the Figure.\label{double}}
\end{figure}

\begin{figure}[htb]
\begin{center}\includegraphics[%
  width=14cm,
  height=4cm]{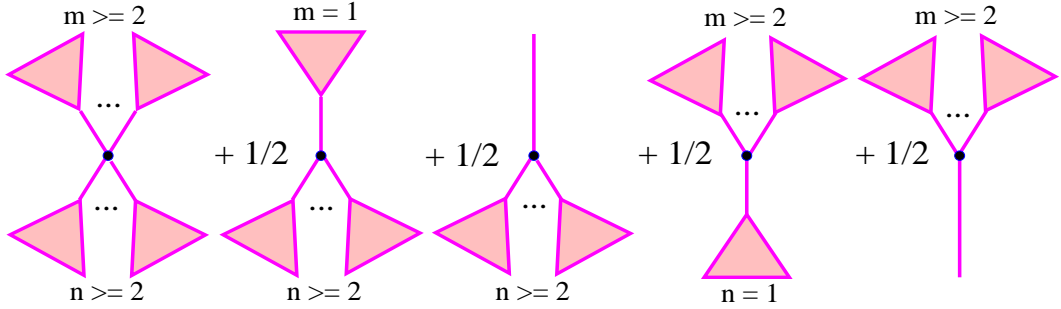}\end{center}
\vspace{-4mm}

\caption{Complete set of {}``tree'' graphs.\label{full-tree}}
\end{figure}

Thus, a general representation for {}``tree'' graphs corresponds
to the diagram with one {}``central'' vertex connected to any number
$m$ of projectile {}``fans'' $\chi_{a}^{{\rm fan}}$ and to any number
$n$ of target ones $\chi_{d}^{{\rm fan}}$; $m,n\geq1$, $m+n\geq3$.
However, here one should be careful with counting some contributions
twice. For example, comparing the subsamples with $m=2$, $n=1$ and
$m=1$, $n=2$, we can see that the graphs of the type of Fig.~\ref{net}b
are present in both cases. Let us consider the first case and express
the lower (target) {}``fan'' using the relation of Fig.~\ref{ffan},
as shown in Fig.~\ref{double}.%
 A similar procedure can be performed in the second case ($m=1$,
$n=2$), which  results in the graphs of Fig.~\ref{double} being
reversed upside-down, with the second diagram in the r.h.s.~staying
unchanged. Thus, correcting for such double counts, the overall {}``tree''
type contribution is given by the set of graphs of Fig.~\ref{full-tree},%
with
\begin{eqnarray}
\chi_{ad}^{{\rm tree}}(s,b)=
\frac{G}{C_{\pi}^{2}}\int_{0}^{Y}\!\! dy_{1}\int\!\! d^{2}b_{1}\;
\left\{ \left(1-e^{-\chi_{a}^{{\rm fan}}}-
\chi_{a}^{{\rm fan}}\right)\left(1-e^{-\chi_{d}^{{\rm fan}}}
-\chi_{d}^{{\rm fan}}\right)\right.\nonumber \\
+\left.\frac{1}{2}(\chi_{a}^{{\rm fan}}+
\chi_{a\pi}^{\mathbb{P}})\left(1-e^{-\chi_{d}^{{\rm fan}}}-
\chi_{d}^{{\rm fan}}\right)
+\frac{1}{2}(\chi_{d}^{{\rm fan}}+\chi_{d\pi}^{\mathbb{P}})\left(1-e^{-\chi_{a}^{{\rm fan}}}-\chi_{a}^{{\rm fan}}\right)\right\} \label{tree}\end{eqnarray}
Here 
 $\chi_{a\pi}^{\mathbb{P}}
 =\chi_{a\pi}^{\mathbb{P}}(s_0\,e^{Y-y_{1}},|\vec{b}-\vec{b}_{1}|)$,
$\chi_{a}^{{\rm fan}}=\chi_{a}^{{\rm fan}}(Y-y_{1},|\vec{b}-\vec{b}_{1}|)$,
$\chi_{d\pi}^{\mathbb{P}}=\chi_{d\pi}^{\mathbb{P}}(s_{0}\,e^{y_{1}},b_{1})$,
$\chi_{d}^{{\rm fan}}=\chi_{d}^{{\rm fan}}(y_{1},b_{1})$.

\begin{figure}[htb]
\begin{center}\includegraphics[%
  width=14cm,
  height=4.5cm]{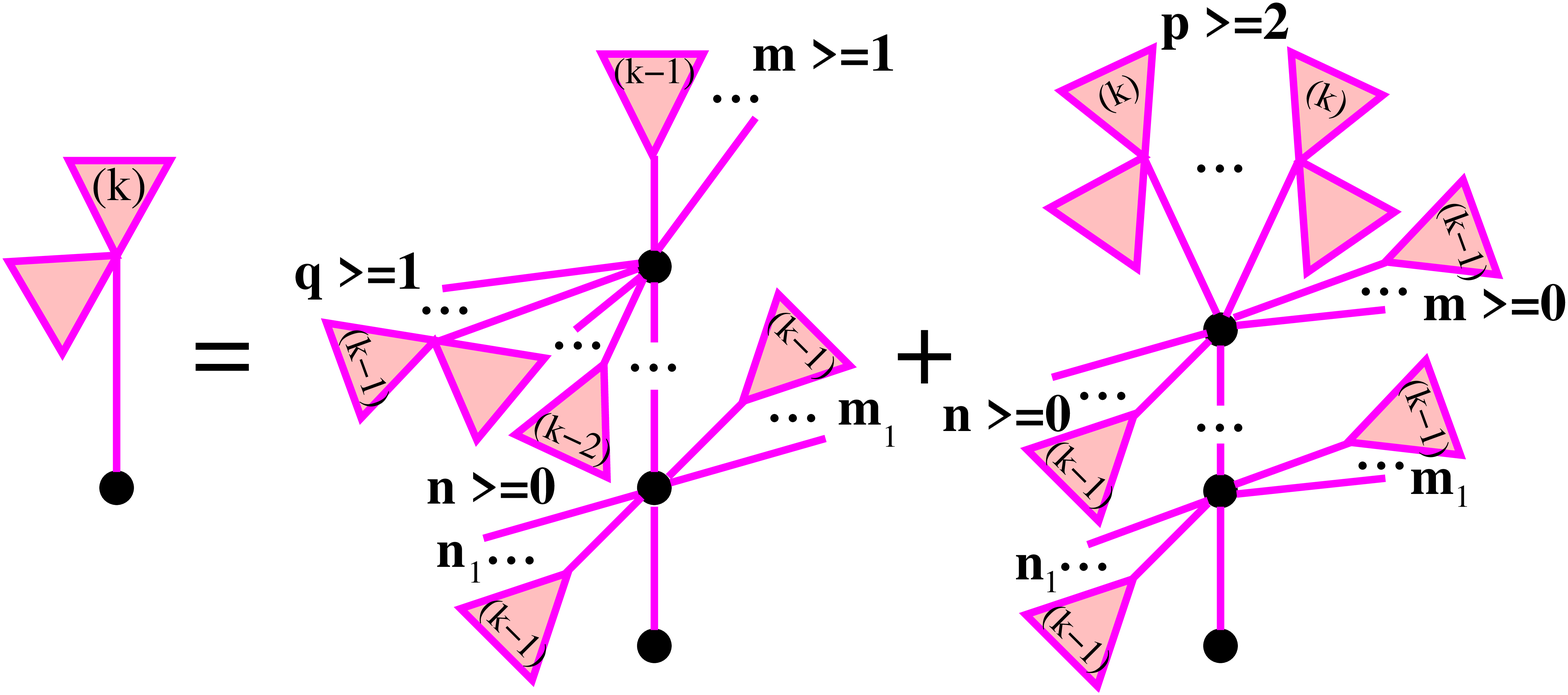}\end{center}

\vspace{-4mm}

\caption{{}``Zig-zag fan'' $\chi_{a|d}^{{\rm zz}(k)}(y_{1},\vec{b}_{1}|Y,\vec{b})$
can be represented as shown in the Figure. \label{zig-rec}}
\end{figure}

Let us come to more general diagrams of the type of Fig.~\ref{net}c,
corresponding to arbitrary {}``nets'' of Pomerons. It is convenient
to introduce some new building blocks. We define the {}``net fan''
of $k$-th order via the recursive equation of Fig.~\ref{freve}:%
\begin{figure}[t]
\begin{center}\includegraphics[%
  width=12cm,
  height=4cm]{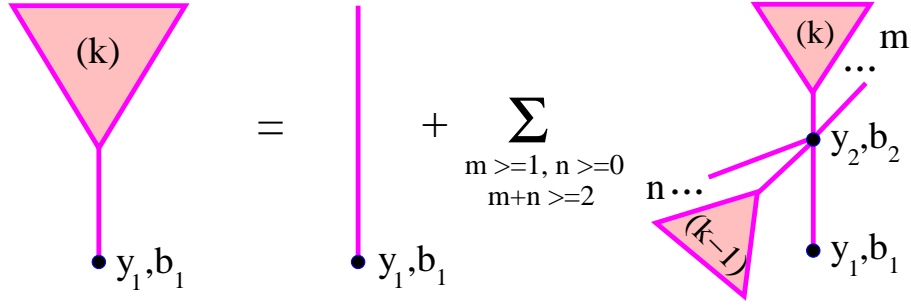}\end{center}

\vspace{-4mm}

\caption{Recursive equation for the {}``net fan'' $\chi_{a|d}^{{\rm net}(k)}(y_{1},\vec{b}_{1}|Y,\vec{b})$.
The vertex $(y_{2},b_{2})$ couples together $m$ projectile {}``net
fans'' of $k$-th order and $n$ target {}``net fans'' of $(k-1)$-th
order.\label{freve}}
\end{figure}
\begin{eqnarray}
\chi_{a|d}^{{\rm net}(k)}(y_{1},\vec{b}_{1}|Y,\vec{b})
=\chi_{a\pi}^{\mathbb{P}}(s_{0}\,e^{y_{1}},b_{1})+\frac{G}{C_{\pi}^{2}}
\int_{0}^{y_{1}}\!\! dy_{2}\int\!\! d^{2}b_{2}\;\;
\chi_{\pi\pi}^{\mathbb{P}}(s_{0}\,e^{y_{1}-y_{2}},|\vec{b}_{1}-\vec{b}_{2}|)
\nonumber \\ \times
\left\{ \left[1-e^{-\chi_{a|d}^{{\rm net}(k)}\!(y_{2},\vec{b}_{2}|Y,\vec{b})}
\right]\:\exp\!
\left(-\chi_{d|a}^{{\rm net}(k-1)}(Y-y_{2},\vec{b}-\vec{b}_{2}|Y,\vec{b})
\right)
-\chi_{a|d}^{{\rm net}(k)}(y_{2},\vec{b}_{2}|Y,\vec{b})\right\} 
\label{g-fan}\end{eqnarray}
Here we set $\chi_{a|d}^{{\rm net}(1)}(y_{1},\vec{b}_{1}|Y,\vec{b})\equiv\chi_{a}^{{\rm fan}}(y_{1},b_{1})$,
$\chi_{a|d}^{{\rm net}(0)}(y_{1},\vec{b}_{1}|Y,\vec{b})\equiv0$.
 By construction, the {}``net fan'' of $k$-th order contains contributions
with a sequence of up to $k$ Pomerons connected to each other in
a {}``zig-zag'' way, such that Pomeron end rapidities are arranged
as $y_{1}>y_{2}<y_{3}>...<y_{k}$. For example, the part of the graph
of Fig.~\ref{net}c, positioned to the left of the vertex $(y_{1},b_{1})$,
can be considered as belonging to the projectile {}``net fan'' of
3rd order; the {}``zig-zag'' is formed by three Pomerons: those
exchanged between the vertices $(y_{1},b_{1})$ and $(y_{2},b_{2})$,
$(y_{2},b_{2})$ and $(y_{3},b_{3})$, and by the leftmost Pomeron
connected to the projectile. Correspondingly, we define a {}``zig-zag
fan'' of k-th order as the difference between $k$-th and $(k-1)$-th
{}``net fans'':
\begin{equation}
\chi_{a|d}^{{\rm zz}(k)}(y_{1},\vec{b}_{1}|Y,\vec{b})
\equiv\chi_{a|d}^{{\rm net}(k)}(y_{1},\vec{b}_{1}|Y,\vec{b})
-\chi_{a|d}^{{\rm net}(k-1)}(y_{1},\vec{b}_{1}|Y,\vec{b})\label{zig-zag}
\end{equation}
Using the representation of Fig.~\ref{freve} for the {}``net fan''
contributions, $\chi_{a|d}^{{\rm zz}(k)}$ can be expressed as shown
in Fig.~\ref{zig-rec}.%
  The two graphs in the r.h.s.~of the Figure differ by their uppermost
vertices: in the 1st graph it couples together $m\geq1$ projectile
{}``net fans'' of $(k-1)$-th order, $q\geq1$ target {}``zig-zag
fans'' of $(k-1)$-th order and any number $n\geq0$ of target {}``net
fans'' of $(k-2)$-th order; the uppermost vertex of the 2nd graph
is coupled to $p\geq2$ projectile {}``zig-zag fans'' of $k$-th
order and to any number $m\geq0$ of projectile and $n\geq0$ of target
{}``net fans'' of $(k-1)$-th order. In addition, both graphs may
contain any number $l\geq0$ of intermediate vertices, coupled correspondingly
to $m_{i}$ projectile and $n_{i}$ target {}``net fans'' of $(k-1)$-th
order; $m_{i},n_{i}\geq0$, $m_{i}+n_{i}\geq1$, $i=1,...,l$.

\begin{figure}[htb]
\begin{center}\includegraphics[%
  width=14.5cm,
  height=4cm]{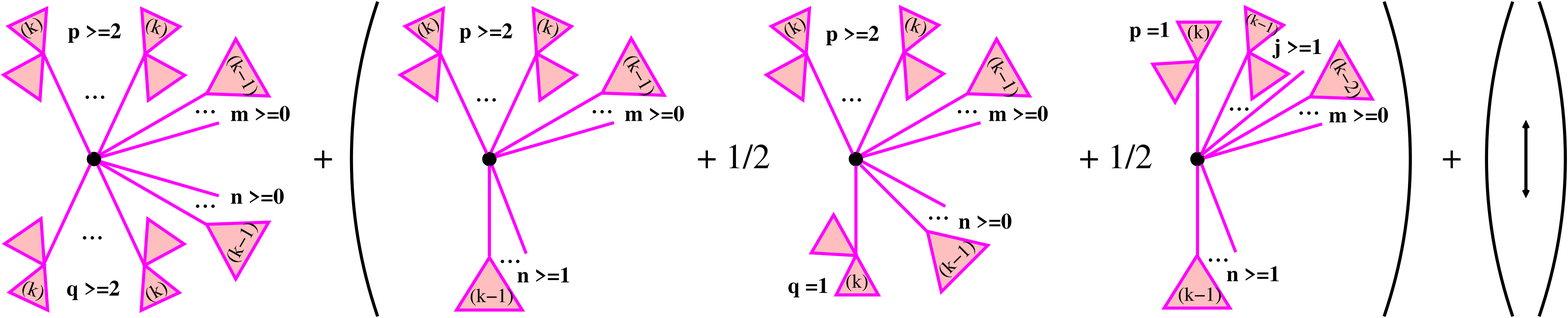}\end{center}

\vspace{-4mm}

\caption{Complete set of enhanced diagrams containing \char`\"{}zig-zag fans\char`\"{}
of $k$-th order.\label{enh-k}}
\end{figure}

Now we can re-sum enhanced diagrams which contain {}``zig-zag fans''
of $k$-th order, starting from $k=2$, using the representation of
Fig.~\ref{zig-rec} to correct for double counts in the same way
as we did for {}``tree'' graphs.
 This results in the set of diagrams of Fig.~\ref{enh-k};%
the corresponding
contribution to quasi-eikonal is
\begin{eqnarray}
\chi_{ad}^{{\rm enh}(k)}(s,b)=
\frac{G}{C_{\pi}^{2}}\int_{0}^{Y}\!\! dy_{1}\int\!\! d^{2}b_{1}\;
\left\{ \!\left(1-e^{-\chi_{a|d}^{{\rm zz}(k)}}-
\chi_{a|d}^{{\rm zz}(k)}\right)e^{-\chi_{a|d}^{{\rm net}(k-1)}}
\right.\nonumber \\
\times\left(1-e^{-\chi_{d|a}^{{\rm zz}(k)}}-
\chi_{d|a}^{{\rm zz}(k)}\right)\,e^{-\chi_{d|a}^{{\rm net}(k-1)}}
\nonumber \\
+\left[\left(1-e^{-\chi_{a|d}^{{\rm zz}(k)}}-\chi_{a|d}^{{\rm zz}(k)}\right)\,
e^{-\chi_{a|d}^{{\rm net}(k-1)}}
\left(1-e^{-\chi_{d|a}^{{\rm net}(k-1)}}+
\frac{1}{2}\chi_{d|a}^{{\rm zz}(k)}\,
e^{-\chi_{d|a}^{{\rm net}(k-1)}}\right)\right.
\nonumber \\
+\left.\left.\frac{1}{2}
\chi_{a|d}^{{\rm zz}(k)}\,\left(e^{-\chi_{a|d}^{{\rm zz}(k-1)}}
-1\right)\,e^{-\chi_{a|d}^{{\rm net}(k-2)}}\,
\left(1-e^{-\chi_{d|a}^{{\rm net}(k-1)}}\right)\right]+
\left[a\leftrightarrow d\right]\right\} \label{delta-k}\end{eqnarray}
Here $\chi_{a|d}^{{\rm X}(k)}=\chi_{a|d}^{{\rm X}(k)}(Y-y_{1},\vec{b}-\vec{b}_{1}|Y,\vec{b})$,
$\chi_{d|a}^{{\rm X}(k)}=\chi_{d|a}^{{\rm X}(k)}(y_{1},\vec{b}_{1}|Y,\vec{b})$,
${\rm X="zz","net"}$.

\begin{figure}[htb]
\begin{center}\includegraphics[%
  width=14.5cm,
  height=4cm]{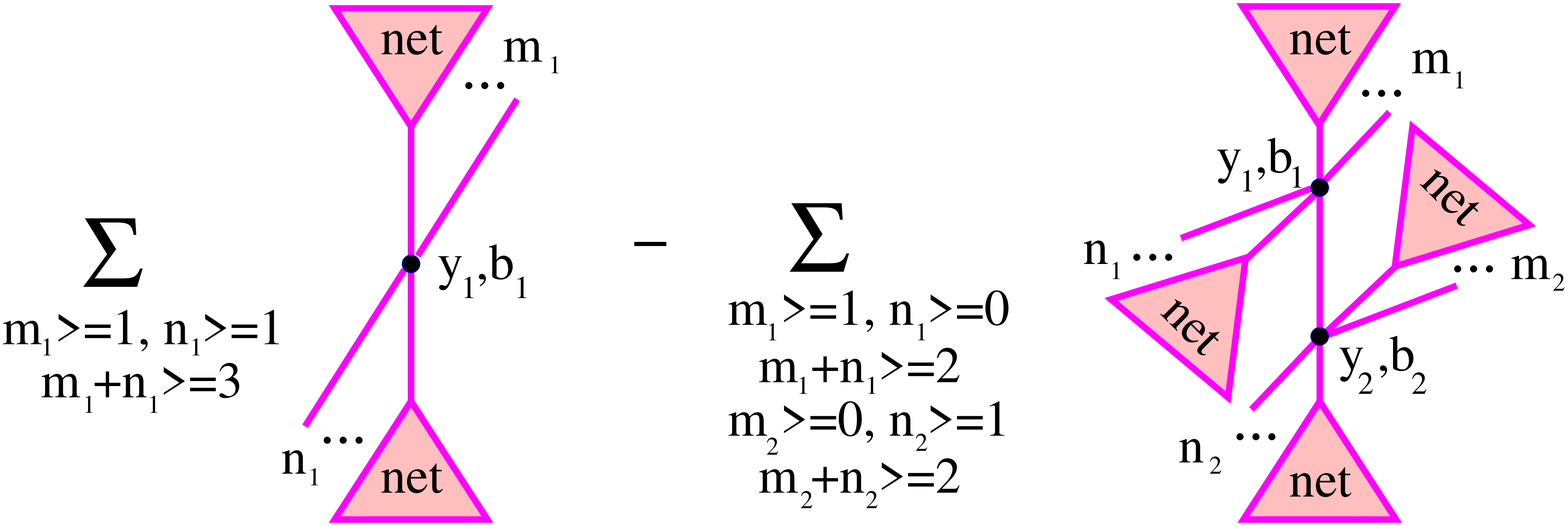}\end{center}

\vspace{-4mm}

\caption{Full set of non-loop diagrams.\label{enh-full}}
\end{figure}

Finally, combining together the contributions $\chi_{ad}^{{\rm tree}}(s,b)$
and $\chi_{ad}^{{\rm enh}(k)}(s,b)$ for any $k\geq2$ and using 
(\ref{g-fan}-\ref{zig-zag}),
we can obtain for the full set of non-loop enhanced diagrams the representation
of Fig.~\ref{enh-full},%
 with the eikonal contribution \begin{eqnarray}
\chi_{ad}^{{\rm enh}}(s,b)=\chi_{ad}^{{\rm tree}}(s,b)
+\sum_{k=2}^{\infty}\chi_{ad}^{{\rm enh}(k)}(s,b)
=\frac{G}{C_{\pi}^{2}}\int_{0}^{Y}\!\! dy_{1}\!\int\!\! d^{2}b_{1}\;
\left\{ \left[\left(1-e^{-\chi_{a|d}^{{\rm net}}(1)}\right)
\right.\right.
\nonumber \\
\times\left.\left(1-e^{-\chi_{d|a}^{{\rm net}}(1)}\right)
-\chi_{a|d}^{{\rm net}}(1)\;\chi_{d|a}^{{\rm net}}(1)\right]
-\frac{G}{C_{\pi}^{2}}\int_{0}^{y_{1}}\!\! dy_{2}\!\int\!\! d^{2}b_{2}\;\;
\chi_{\pi\pi}^{\mathbb{P}}(s_{0}\,e^{y_{1}-y_{2}},|\vec{b}_{1}-\vec{b}_{2}|)
\nonumber \\
\times\left.\left[\left(1-e^{-\chi_{a|d}^{{\rm net}}(1)}\right)\,
e^{-\chi_{d|a}^{{\rm net}}(1)}-\chi_{a|d}^{{\rm net}}(1)\right]\,
\left[\left(1-e^{-\chi_{d|a}^{{\rm net}}(2)}\right)\,
e^{-\chi_{a|d}^{{\rm net}}(2)}-\chi_{d|a}^{{\rm net}}(2)\right]\right\}
 \label{delta-tot}\end{eqnarray}
Here we used the abbreviations $\chi_{a|d}^{{\rm net}}(i)=\chi_{a|d}^{{\rm net}}(Y-y_{i},\vec{b}-\vec{b}_{i}|Y,\vec{b})$,
$\chi_{d|a}^{{\rm net}}(i)=\chi_{d|a}^{{\rm net}}(y_{i},\vec{b}_{i}|Y,\vec{b})$,
$i=1,2,$ and introduced general {}``net fan'' contribution as
 $\chi_{a|d}^{{\rm net}}=\lim_{k\rightarrow\infty}\chi_{a|d}^{{\rm net}(k)}$.
Using (\ref{g-fan}), we obtain for the latter the recursive equation
\begin{eqnarray}
\chi_{a|d}^{{\rm net}}(y_{1},\vec{b}_{1}|Y,\vec{b})=
\chi_{a\pi}^{\mathbb{P}}(s_{0}\,e^{y_{1}},b_{1})+\frac{G}{C_{\pi}^{2}}
\int_{0}^{y_{1}}\!\! dy_{2}\int\!\! d^{2}b_{2}\;\;
\chi_{\pi\pi}^{\mathbb{P}}(s_{0}\,e^{y_{1}-y_{2}},|\vec{b}_{1}-\vec{b}_{2}|)
\nonumber \\
\left\{ \left[1-e^{-\chi_{a|d}^{{\rm net}}(y_{2},\vec{b}_{2}|Y,\vec{b})}\right]
\,\exp\!\left(-\chi_{d|a}^{{\rm net}}(Y-y_{2},\vec{b}-\vec{b}_{2}|Y,\vec{b})\right)
-\chi_{a|d}^{{\rm net}}\!\!\left(y_{2},\vec{b}_{2}|Y,\vec{b}\right)\right\} 
\label{net-fan}\end{eqnarray}

\section{Numerical results }

\begin{figure}[htb]
\begin{center}\includegraphics[%
  width=7.5cm,
  height=6cm]{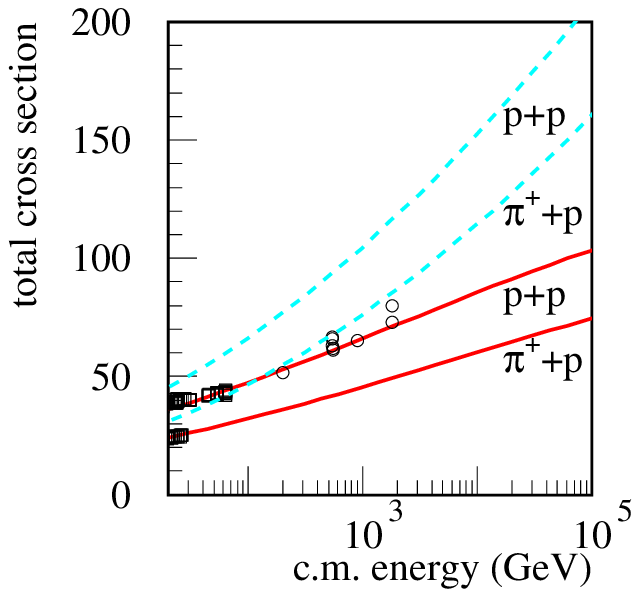}\includegraphics[%
  width=7.5cm,
  height=6cm]{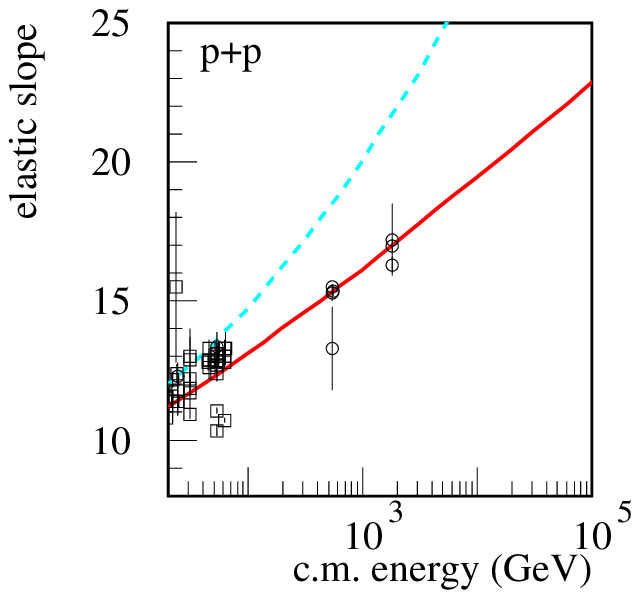}\end{center}

\vspace{-4mm}

\caption{Total cross section (left) and elastic scattering slope (right) as
calculated with and without enhanced contributions - solid and dashed
lines correspondingly.  The compilation of data is from \cite{cas98}.
\label{sigt}}
\end{figure}

The obtained expressions allowed us to calculate hadronic elastic
scattering amplitudes and correspondingly total cross sections and
elastic scattering slopes with enhanced contributions taken
into account. Here $f_{ad}$, $\sigma_{ad}^{{\rm tot}}$, $B_{ad}^{{\rm el}}$
are given by usual expressions (\ref{f_ad}-\ref{b_el}), with the
Pomeron quasi-eikonal $\chi_{ad}^{\mathbb{P}}$ being replaced by
$\chi_{ad}^{{\rm tot}}=\chi_{ad}^{\mathbb{P}}+\chi_{ad}^{{\rm enh}}$.
Technically, the {}``net fan'' contribution $\chi_{a|d}^{{\rm net}}$
has been obtained solving (\ref{net-fan}) iteratively and substituted
to (\ref{delta-tot}) to calculate enhanced diagram contribution $\chi_{ad}^{{\rm enh}}$.
Concerning the parameter choice we used the usual values $C^2_{p}=1.5$,
$C_{\pi}=1.6/C_{p}$,  $\gamma_{\pi}=2/3\gamma_{p}$ \cite{kai82},
and from comparison to data obtained $\alpha_{\mathbb{P}}(0)=1.18$,
$\alpha_{\mathbb{P}}'(0)=0.195$ GeV$^{-2}$, $\gamma_{p}=1.59$ GeV$^{-1}$,
$R_{p}^{2}=1.8$ GeV$^{-2}$, $R_{\pi}^{2}=0.7$ GeV$^{-2}$, 
$G_{3\mathbb{P}}=9\cdot 10^{-3}$ GeV$^{2}$. 
Thus, for the triple-Pomeron coupling
we have $r_{3\mathbb{P}}=4\pi \,G\,C_{\pi}\gamma_{\pi}^3=0.18$
GeV$^{-1}$ compared to  0.12 GeV$^{-1}$ and 0.083  GeV$^{-1}$  in  
 \cite{kai86} and \cite{bon01} correspondingly.
The results for $\sigma_{pp}^{{\rm tot}}$, $\sigma_{\pi p}^{{\rm tot}}$,
$B_{pp}^{{\rm el}}$ are shown in Fig.~\ref{sigt}%
 as calculated with the full scheme or based on the bare Pomeron
eikonal $\chi_{ad}^{\mathbb{P}}$. 
In practice, it is sufficient to take into consideration only 
the "tree" $\chi_{ad}^{{\rm tree}}$ and  the first "zig-zag"
 $\chi_{ad}^{{\rm enh}(2)}$ corrections, 
 i.e.~to use for the enhanced contribution
$\tilde \chi_{ad}^{{\rm enh}}=\chi_{ad}^{{\rm tree}}+\chi_{ad}^{{\rm enh}(2)}$
instead of $\chi_{ad}^{{\rm enh}}$ defined in (\ref{delta-tot});
 the difference for the calculated
cross sections is below percent level. This is because
the contributions $\chi_{ad}^{{\rm enh}(k)}$ for $k\geq 3$ are
 suppressed
by exponential factors in the same way as for  "loop" diagrams in (\ref{dump}).

Let us finally verify that the developed scheme approaches the asymptotic 
result (\ref{asy-tot}) in the \char`\"{}dense\char`\"{} limit.
Indeed, neglecting the radius of multi-Pomeron vertices,
at  $s\rightarrow\infty$, $b\rightarrow0$ and for 
$\alpha_{\mathbb{P}}(0)-4\pi\, G\,\gamma_{\pi}^{2}>1$
 we can  obtain the solution of (\ref{net-fan}) as
$\chi_{a|d}^{{\rm net}}(y_{1},\vec{b}_{1}|Y,\vec{b})\simeq 
\chi_{a\pi}^{\mathbb{P}}(s_{0}\,e^{y_{1}},b_{1})
+\Delta\chi_{a\pi}^{{\rm asymp}}(s_{0}\,e^{y_{1}},b_{1})$,
 $\Delta\chi_{a\pi}^{{\rm asymp}}$ being defined
in  (\ref{asy-tot}). 
 Substituting this to  (\ref{delta-tot}), we see
that the enhanced contribution $\chi_{ad}^{{\rm enh}}$ 
reduces to  the asymptotic form  (\ref{asy-tot}):
$\chi_{ad}^{{\rm enh}}(s,b)\simeq 
\Delta\chi_{ad}^{{\rm asymp}}(s,b)$.

In conclusion, we re-summed dominant enhanced contributions to elastic
hadron-hadron scattering amplitude to all orders. Although the numerical
calculations have been performed using the simple Pomeron exchange
amplitude (\ref{ampl-P}-\ref{lambda-P}), the obtained formulas can
be used for a different functional form of $f_{ad}^{\mathbb{P}}(s,b)$.
In principle, one may apply similar techniques in the perturbative
QCD, using the BFKL Pomeron amplitude \cite{lip86}, provided eikonal
approximation remains applicable for multi-Pomeron vertices.

\bigskip
{\small The author is indebted to R. Engel and A.~B.~Kaidalov for valuable
discussions.}

\end{document}